\documentclass[twocolumn, amssymb, amsmath,nobibnotes,nofootinbib, aps, prl,showpacs,10pt]{revtex4}
\usepackage[dvips]{graphicx}
\begin{document}
\title{Reentrant spin glass state in Mn doped Ni$_2$MnSn shape memory alloy}
\author{S. Chatterjee$^1$}
\author{S. Giri$^1$}
\author{S. K. De$^2$}
\author{S. Majumdar$^1$}
\email{sspsm2@iacs.res.in} 
\affiliation{$^1$Department of Solid State Physics and Center for Advanced Materials,  Indian Association for the Cultivation of Science, 2A \& B Raja S. C. Mullick Road, Jadavpur, Kolkata 700 032, India }
\affiliation{$^2$Department of Materials Science, Indian Association for the Cultivation of Science, 2A \& B Raja S. C. Mullick Road, Jadavpur, Kolkata 700 032, India}
\begin{abstract}
The ground state properties of the ferromagnetic shape memory alloy of nominal composition  Ni$_2$Mn$_{1.36}$Sn$_{0.64}$ have been studied by dc magnetization and ac susceptibility measurements. Like few other Ni-Mn based alloys, this sample  exhibits exchange bias phenomenon. The observed exchange bias pinning was found to originate right from the temperature where a step-like anomaly is present in the zero-field-cooled magnetization data. The ac susceptibility study indicates the onset of spin glass freezing  near this step-like anomaly with clear frequency shift. The sample can be identified as a reentrant spin glass with both ferromagnetic and glassy phases coexisting together at low temperature at least in the field-cooled state. The result provides us an comprehensive view to identify the magnetic character of various Ni-Mn-based shape memory alloys with competing magnetic interactions.   

\end{abstract}
\maketitle

Recently Ni$_2$Mn$_{1+x}$Z$_{1-x}$ (Z = In, Sn, and Sb) based ferromagnetic shape memory alloys (FSMAs)  have attracted considerable attention due to their multifunctional properties, which include magnetic sueprelasticity, giant magnetoresistance, large inverse magnetocaloric effect and magnetic memory effect~\cite{kainuma1, krenke, kainuma2, manosa, chatterjee}. The observed phenomena are primarily related to the magnetic field ($H$) induced reverse transition across the martensitic transformation (MT)~\cite{koyama}. The stoichiometric Heusler compositions (Ni$_2$MnZ) are all ferromagnetic with the Curie point ($T_C$)  lying just above room temperature. The excess Mn doping at the expense of  Z atoms induces structural instability in the system leading to the ferromagnetic shape memory effect. Short range antiferromagnetic (AFM) interaction between the  excess Mn (at the Z site) and the  original  Mn atoms has been predicted  for the doped alloys~\cite{enkovaara, afm}. Although,  predominantly ferromagnetic (FM) character is present in the Mn doped alloys with $T_C$ around  300 K,  the AFM correlation is evident from the gradual decrease of saturation moment with increasing amount of excess Mn~\cite{enkovaara, khan1}. Diffuse peaks observed in the powder neutron diffraction data of Ni-Mn-Sn alloys also indicate the existence of incipient AFM coupling~\cite{brown}.  

\par
Evidently, the magnetic nature of the ground state of the alloys may not be very simple. A fascinating evidence for the complex ground state of the alloys is the recently observed exchange bias (EB)  phenomenon in bulk samples~\cite{khan1, khan2, li}. EB is  referred to the shift of the center of the magnetic hysteresis loop from the origin when the sample has been cooled from high temperature in presence of magnetic field~\cite{eb}. The origin of EB is generally ascribed to the presence of FM and AFM interfacial coupling in a heterogeneous sample. EB has also been observed in materials having FM/spin-glass (SG) and  FM/Ferrimagnet interfaces ~\cite{ali,patra} other than FM/AFM systems. However in all the cases,  it is required that the ordering temperature ($T_{NF}$) for the non-ferromagnetic phase (may be AFM, SG or ferrimagnet) should be lower than the ferromagnetic Curie point ($T_C$).  An interesting characteristics of EB in  all Ni$_2$Mn$_{1+x}$Z$_{1-x}$ alloys is that it vanishes above a certain temperature $T_B$,  which  coincides with the foot of a step-like anomaly observed in the zero-field-cooled magnetization versus temperature  data. In the perspective of EB, such a temperature is generally referred as the exchange bias blocking temperature with a value $T_B \leq T_{NF}$.

\par
Although there are several reports on the observation of EB in  Heusler based shape memory alloys, very little efforts have been given to understand the origin of the effect. Undoubtedly, EB confirms the presence of a non-ferromagnetic phase in the otherwise FM alloy. In order to understand  the  nature of the ground state  and the significance of $T_B$, we have investigated the magnetic properties of  one of the  alloys (Ni$_2$Mn$_{1.36}$Sn$_{0.64}$) showing EB. A reentrant spin glass type behavior  with freezing temperature ($T_f$) above $T_B$ is evident from our study.

\par 
The polycrystalline samples  were prepared by argon arc melting the constituent elements. The dc magnetization ($M$) and the ac susceptibility ($\chi_{ac}$) of the samples were  measured by Quantum Design SQUID magnetometer (MPMS 6, Ever-cool model) and commercial cryogen free vibrating sample magnetometer from Cryogenic Ltd., UK in the temperature($T$) range 5-300 K.

%%%%%%%%%%%%%%%%%%%%%%%%FIGURE 1%%%%%%%%%%%%%%%%%%%%%%%%
\begin{figure}[t]
\begin{center}
\includegraphics[width = 7 cm]{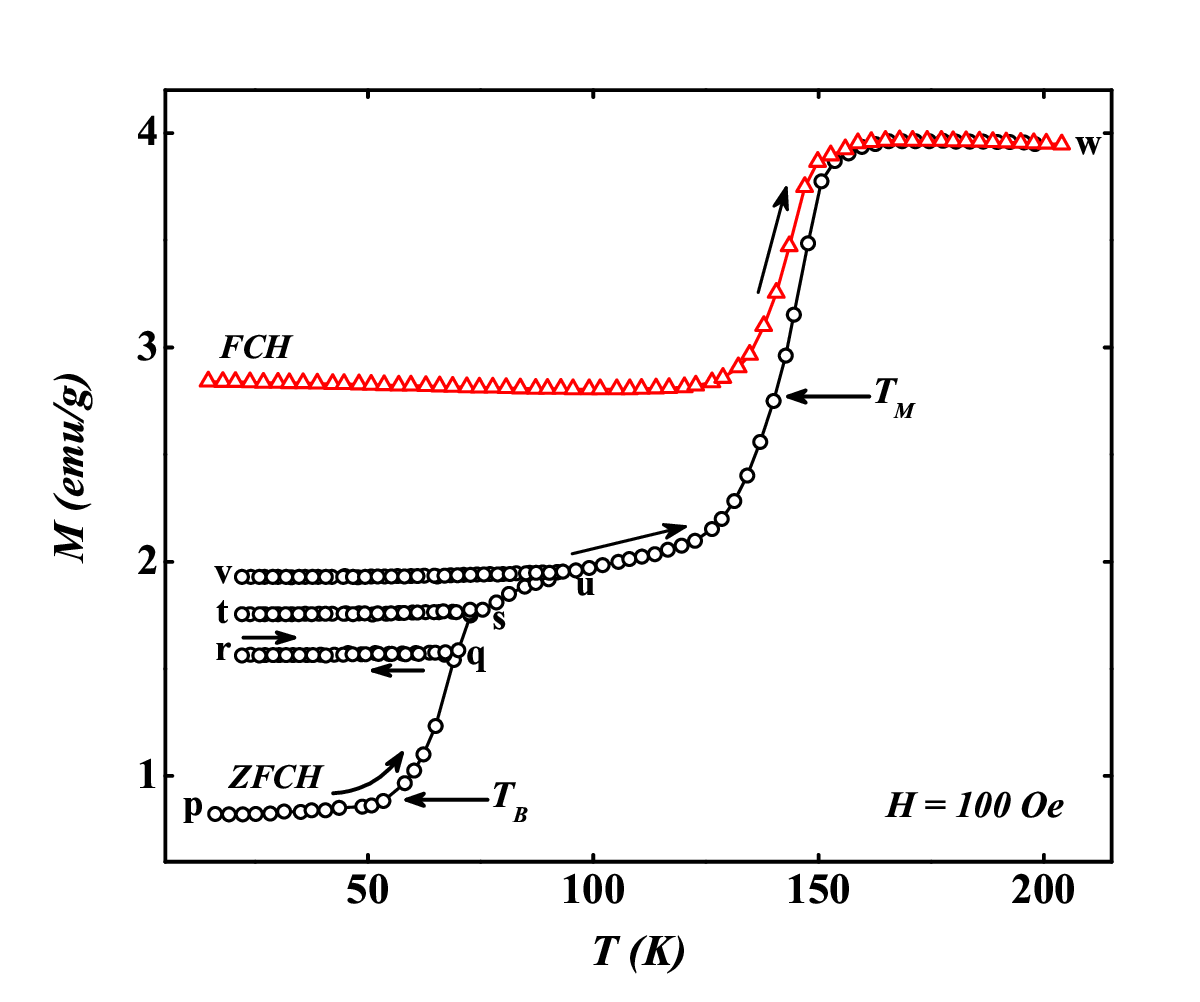}
\caption {Magnetization is plotted as a function of temperature in zero-field-cooled heating (ZFCH) and field-cooled heating (FCH) conditions in presence of 100 Oe of external magnetic field. The martensitic tranformation and the EB blocking temperatures are indicated by $T_M$ and $T_B$ in the figure.  Minor temperature cycling from a zero-field-cooled state above $T_B$ are  also shown in the figure (see text for details).}
\end{center}
\end{figure}
%%%%%%%%%%%%%%%%%%%%%%%%%%%%%%%%%%%%%%%%%%%%%%%%%%%%%%%%

Figure 1 shows the $M$ versus $T$ data in zero-field-cooled heating (ZFCH) and field-cooled heating (FCH) sequences in presence of 100 Oe of applied magnetic field. The anomaly in the $M$($T$) data around 140 K  indicates the signature of MT~\cite{chatterjee1}, which has been indicated by $T_M$.  The  step-like anomaly in the ZFCH data just below 80 K is observed in the sample (the foot  of which is indicated by $T_B$), which is the typical signature of EB blocking temperature  reported in case of various Ni$_2$Mn$_{1+x}$Z$_{1-x}$ alloys.

\par
In order to ascertain the true magnetic character of the  anomaly above $T_B$, we have recorded some minor heating-cooling lines around the step. In this protocol, the sample was first zero field cooled down to 20 K. Then the magnetization was measured in presence of 100 Oe field by varying the sample temperature back and forth from different turning points ({\it e.g.} q, s, u). In other words, sample temperature was increased from 20 K  to a certain temperature $T_m$ (70 K $\leq T_m \leq$ 95 K) and cooled back to 20 K. It was repeated for several  values of $T_m$. These minor lines (qr-rq, st-ts, uv-vu etc.) do not follow the master ZFCH curve (pqsu), rather they trace different curves depending on the  magnetization value at the turning point. Each individual minor curve also do not show any thermal hysteresis, which one should not also expect as we are well below  the first order MT.  However, the present measurement  indicates the signature of strong thermo-magnetic irreversibility which exists locally around the step-like anomaly, and it might be the  onset of some glassy magnetic phase in the system.

%%%%%%%%%%%%%%%%%%%%%%%%FIGURE 2%%%%%%%%%%%%%%%%%%%%%%%%
\begin{figure}[t]
\begin{center}
\includegraphics[width = 7 cm]{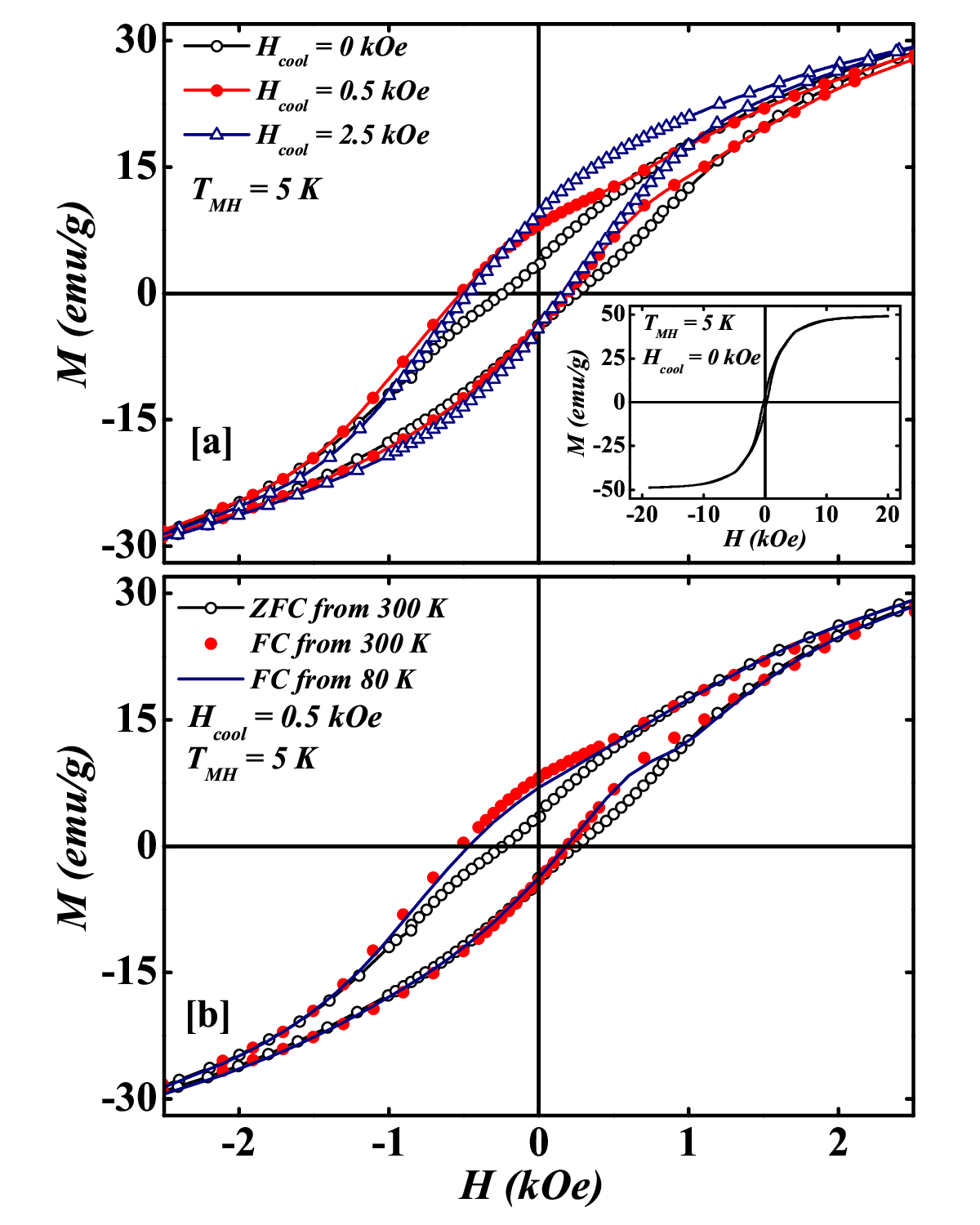}
\caption {(a) Isothermal magnetization loops at 5 K after the sample being cooled in $H$ = 0, 0.5 and 2.5 kOe from 300 K. (b) Isothermal magnetization loops at 5 K after the sample being cooled in zero-field,  in 0.5 kOe from 300 K, and in 0.5 kOe from 80 K. The magnetization loops were recorded by varying the field between $\pm$20 kOe, while  data is shown here between  $\pm$ 2.5 kOe for clarity. A full ($\pm$ 20 kOe) loop on the zero-field-cooled state is shown in the inset of (a).}  
\end{center}
\end{figure}
%%%%%%%%%%%%%%%%%%%%%%%%%%%%%%%%%%%%%%%%%%%%%%%%%%%%%%%%

\par
Now let us look at the EB behavior of the sample with respect to the different cooling fields ($H_{cool}$) and $M-H$ hysteresis loop temperatures ($T_{MH}$). In the previous reports on EB, the $M-H$ isotherms on the field-cooled and the zero-field-cooled states were recorded by cooling the sample from room temperature. Figure 2 (a) shows such isotherms where sample was cooled from 300 K at different fields ($H_{cool}$ = 0, 0.5, 2.5 kOe are only shown here for clarity).  The zero-field-cooled loop is perfectly symmetric with a rather soft FM character (coercive field $\sim$ 250 Oe), while field cooling makes it  asymmetric with shift of the loop both in the field and the magnetization axes. The shift in the horizontal field axis ($H_E$)  and the vertical magnetization axis ($M_E$) are plotted in fig. 3 (a) and (b)  as a function of $H_{cool}$  and  $T_{MH}$ respectively.  $H_E$ increases with $H_{cool}$ sharply at low field, attains a maximum for $H_{cool} \approx$  0.5 kOe and beyond that it decreases slowly with increasing $H_{cool}$. A considerable value of $H_E$ (45 Oe) is observed for very low value of the cooling field ($H_{cool}$ = 75 Oe). On the other hand,  EB   decreases with increasing $T_{MH}$,  and it almost vanishes at $T_{MH}$ = 50 K, which coincides with  $T_B$,  as defined in fig. 1. This behavior is in line with the previous report of EB in different Ni-Mn-Z samples.

%%%%%%%%%%%%%%%%%%%%%%%%FIGURE 3%%%%%%%%%%%%%%%%%%%%%%%%%
\begin{figure}[t]
\begin{center}
\includegraphics[width = 7 cm]{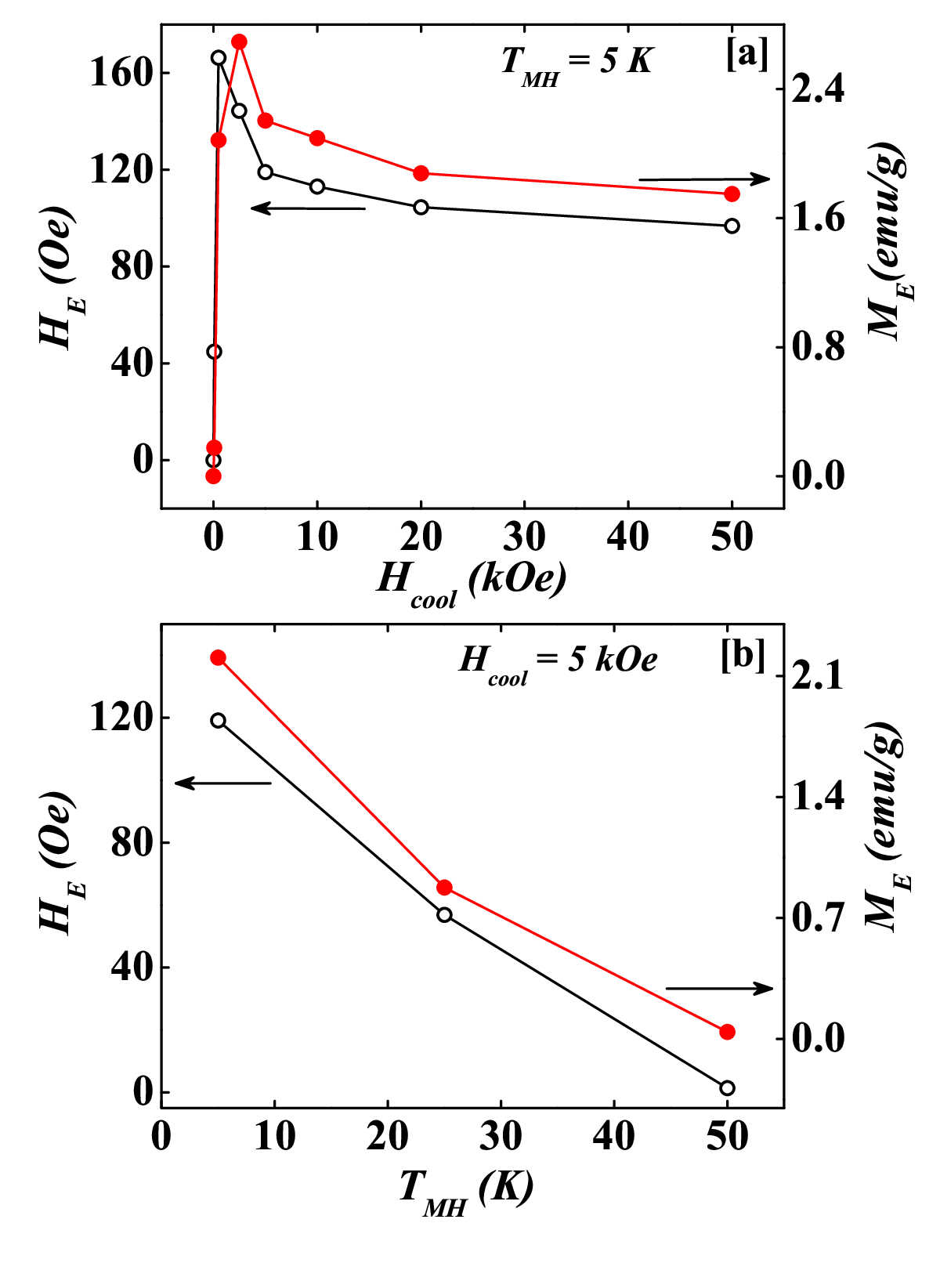}
\caption { Exchange field ($H_E$)  and shift in magnetization ($M_E$) are plotted against (a) the cooling field and  (b) against the temperature of isothermal magnetization measurement.}
\end{center}
\end{figure}
%%%%%%%%%%%%%%%%%%%%%%%%%%%%%%%%%%%%%%%%%%%%%%%%%%%%%%%%
 
\par
The most interesting observation is the $T_{cool}$ (the temperature from where the sample is field cooled) dependence of the EB. We have recorded EB by field cooling the sample from (i) 300 K, (ii) just above the step-like anomaly (80 K) and (iii) from below the step-like anomaly (40 K). In case of (ii) and (iii), the  sample was zero-field-cooled from 300 K to 80 K and 40 K respectively and then it was filed-cooled down to  5 K.  There is practically no difference between the EB behavior for data taken by field-cooling from 300 K  and 80 K. This has been  shown in fig. 2 (b), where solid line and the filled circles loops almost coincide. Evidently, both the protocols show identical values of $H_E$. However when the sample was cooled from 40 K, no asymmetry in the $M-H$ loop  was observed indicating $H_E = $ 0. The exchange pinning, therefore,  actually takes place due to field cooling through the step-like anomaly, which is well below the martensitic transformation temperature. The  anomaly  signifies the second non-ferromagnetic ordering below the $T_C$ and it is responsible for the EB in the sample.

%%%%%%%%%%%%%%%%%%%%%%%%FIGURE 4%%%%%%%%%%%%%%%%%%%%%%%%
\begin{figure}[t]
\begin{center}
\includegraphics[width = 6 cm]{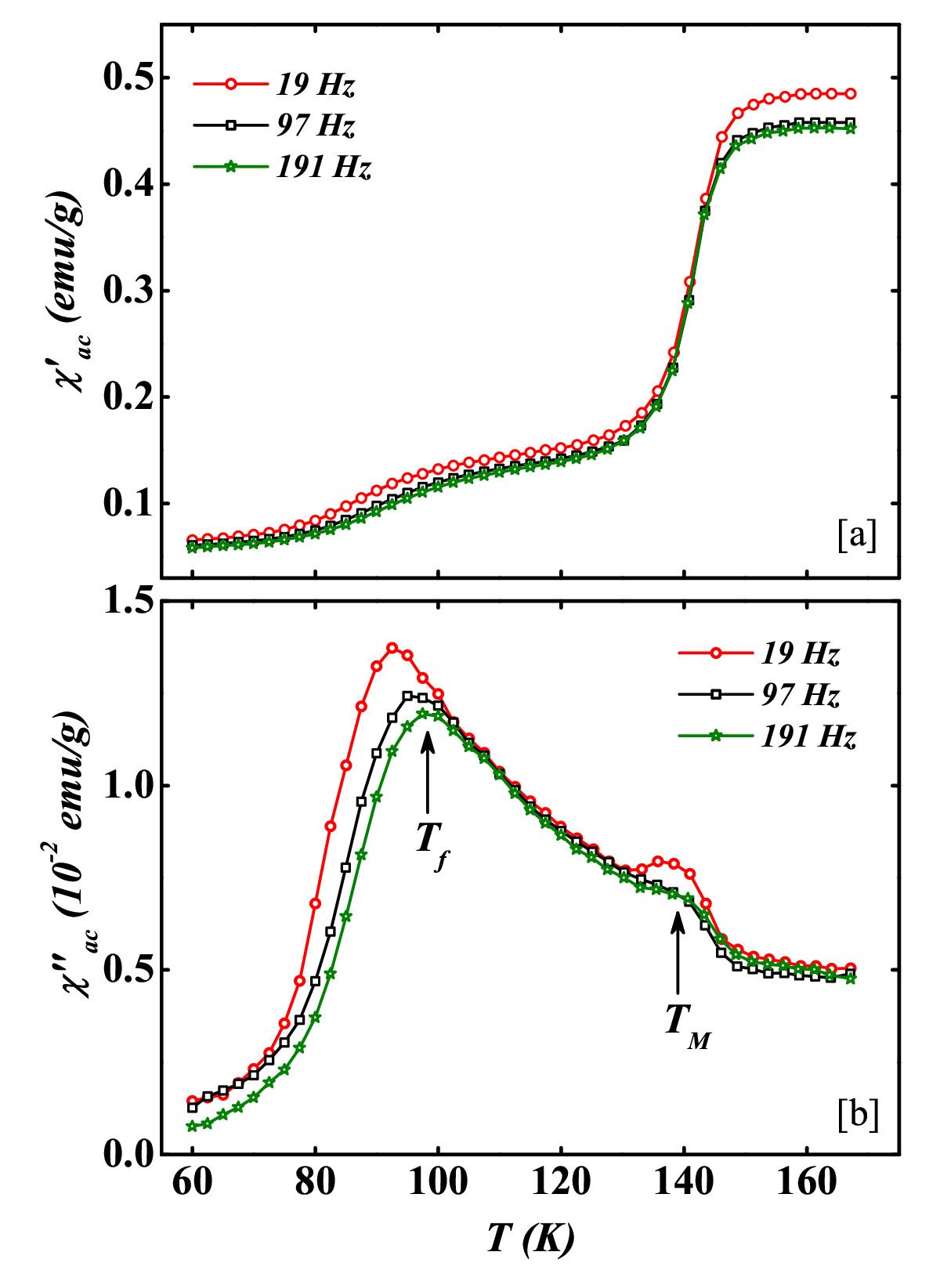}
\caption {(a) Real  part and (b) imaginary part of the ac susceptibility measured at different applied frequencies of the ac signal. The peak magnitude of the applied ac field is 10 Oe.}
\end{center}
\end{figure}
%%%%%%%%%%%%%%%%%%%%%%%%%%%%%%%%%%%%%%%%%%%%%%%%%%%%%%%%

%%%%%%%%%%%%%%%%%%%%%%%%FIGURE 5%%%%%%%%%%%%%%%%%%%%%%%%
\begin{figure}[t]
\begin{center}
\includegraphics[width = 8 cm]{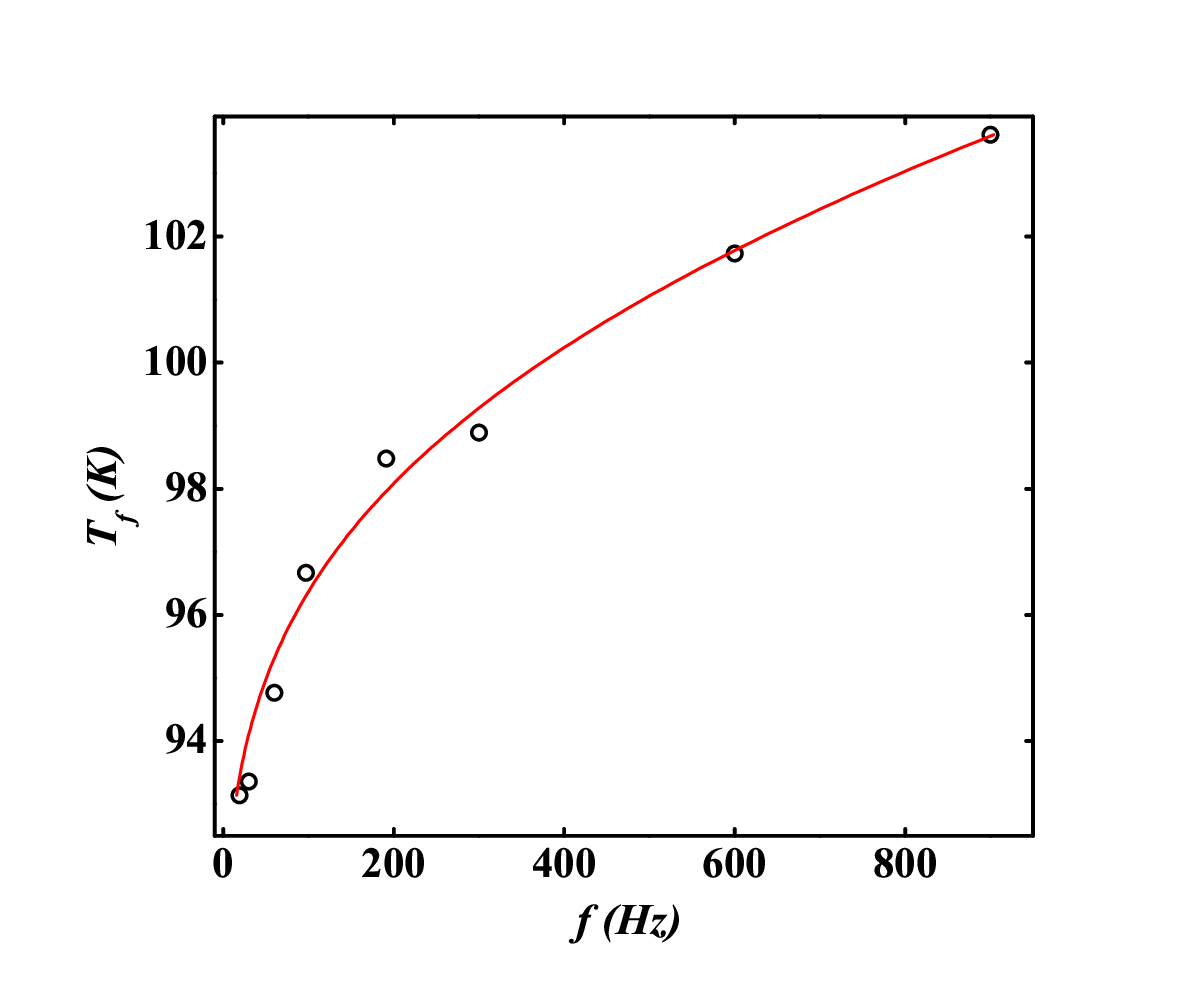}
\caption {Frequency dependence of the peak in the imaginary part of the ac susceptibility observed near the step-like anomaly. The solid line is  the Vogel-Fulcher fit to the data.}
\end{center}
\end{figure}
%%%%%%%%%%%%%%%%%%%%%%%%%%%%%%%%%%%%%%%%%%%%%%%%%%%%%%%%

Considering the fact that thermomagnetic irreversibility  exists near the step-like anomaly, we have performed ac susceptibility measurement at different frequencies  to ascertain the existence of any  spin glass like state. The ac susceptibility data are depicted in the temperature range 60 to 165 K, which  covers the martensitic transformation temperature (indicated by $T_M$ in fig. 4 (b)) and the step-like anomaly (above $T_B$ in fig. 1). Clear feature is  visible in the real part ($\chi_{ac}'$) as well as the out of phase part ($\chi_{ac}''$) of the ac susceptibility data near the step-like anomaly. However, it is extremely prominent in $\chi_{ac}''$, where a broad peak is present. Strong feature in the out of phase component is  an indication of the onset of glassyness of the system.  Interestingly, the feature near the  step-like anomaly shows strong frequency ($f$) dependence in both $\chi_{ac}'$ and $\chi_{ac}''$ data. This is  the indication  of  a glassy magnetic state at low temperature. Expectedly, the  first order MT at $T_M$ does not show any such frequency dependence and its signature in $\chi_{ac}''$  is rather weak. Therefore, the frequency shift of the step-like anomaly is not directly linked to the metastability 
arising from the first order MT.  Figure 5 shows the frequency dependence of the peak temperature in $\chi_{ac}''$  near the step-like anomaly (denoted by $T_f$ in fig. 4 (b)). The relative frequency shift is often expressed  as $\mathcal{P}$ = $\Delta T_f/T_f (\Delta log_{10}\omega)$, where $\omega$ is the angular frequency ($\omega = 2\pi f$). For the present sample, $\mathcal{P}$ was found to be 0.06, which lies well within the region of values  for canonical spin glasses~\cite{mydosh}.  For spin glasses,  empirical Vogel-Fulcher law, $\omega = \omega_0 \exp[-E_a/k_B(T_f - T_0)]$, is widely used to analyze the frequency dependence of the spin freezing temperature ($T_f$)~\cite{mydosh, binder}. Here $E_a$ is the activation energy of the spin glass, $T_0$ is the Vogel-Fulcher temperature, and $\omega_0$ is the characteristic frequency for spin freezing.  In our case freezing temperature $T_f (\omega)$ is the peak temperature of $\chi_{ac}''$, which lies at the onset point of the step-like anomaly in the dc magnetization.  The Vogel-Fulcher fitting of the data has been shown by a solid line in fig. 5. The fitted parameters, $E_a$ $\omega_0$ and $T_0$ was found to be 100.85 K, 5$\times$ 10$^5$ Hz,and  81.5 K respectively. 

\par
From the ac susceptibility analysis, it is clear that the sample show spin-glass like behavior with a frequency dependent glassy transition above $T_B$. This is a remarkable example of glassy magnetic ground state for the ``Ferromagnetic'' shape memory alloy. In the present case, the spin glass is not  of conventional type rather  it has a reentrant character, where the sample first orders ferromagnetically from a paramagnetic state below $T_C$ = 340 K, and below a temperature $T_f$ near the step-like anomaly, the sample further transforms from the ordered FM to a disordered glassy magnetic phase. Reentrant spin glass (RSG) behavior has been observed in variety of magnetic systems including metallic alloys and oxides~\cite{rsg1, rsg2}. Our data on the EB measurement show that low temperature non-ferromagnetic ordering/freezing  in the system responsible for the EB occurs at around the step-like anomaly, and {\it field cooling from this temperature is sufficient enough to obtain the desired field shift} in the isothermal magnetization measurements.  The spin glass-like freezing at $T_f$ is actually responsible for the exchange pinning of the ferromagnetic moments and we can identify $T_{NF}$ to be $T_f$. The MT at $T_M$ does not have a direct role for the observed EB.

\par
The observed RSG phase develops from spin frustration arising from the short range AFM interaction which prevails  between the excess Mn and the original Mn atoms in Ni$_2$Mn$_{1+x}$Z$_{1-x}$ alloys. The AFM interaction gets enhanced in the martensite phase~\cite{afm}, and eventually below $T_f$, the sample enters into an RSG phase. The step-like anomaly in the ZFC data is due to the drop in $M$ arising from the random spin freezing. It is now pertinent  to know the nature of the RSG phase, {\it i.e.} whether the ground state has a true SG character or both FM and SG coexists below $T_f$. The sample shows FM nature at low temperature in the FC state (see fig. 1), and in addition the observed EB for $H_{cool}$ as low as 75 Oe indicates that FM and SG phase indeed coexists at 5 K for sample cooled in  presence of low magnetic field. Coexistence of SG and FM phase is not unusual for RSG~\cite{rsg1}, and  models based on mean field theory predict that FM phase exists below the RSG transition~\cite{parisi}. Notably, in contrary to the conventional spin glasses, the peak magnitude at $T_f$ in $\chi_{ac}''$ decreases with increasing $f$. Similar anomalous change in the peak magnitude has been observed in case of phase separated manganites~\cite{deac}. For the present sample, this might also be the indication of the coexistence of FM and SG phases even at zero applied field.

\par
In conclusion, we observe RSG behavior in the ferromagnetic shape memory alloy of composition Ni$_2$Mn$_{1.36}$Sn$_{0.64}$. The spin glass transition was found to be responsible for the observed EB phenomenon in the alloy.  We would also  like to add that the very similar RSG phase, and associated EB were also observed in the alloy  Ni$_2$Mn$_{1.44}$Sn$_{0.56}$, which has a slightly higher Mn concentration. This second composition also shows frequency dependence in $\chi_{ac}$ near  the step-like anomaly, although $T_f$ has a much higher value (140 K). This prompts us to propose a generic view of the RSG type  magnetic ground state of various Ni$_2$Mn$_{1+x}$Z$_{1-x}$ type FSMAs showing EB, where a common spin freezing phenomena is responsible for the observed exchange bias in all these alloys.

\par

The present work is financially supported by the grant from CSIR, India.

\end{document}